\documentclass[aps,prd,onecolumn,floatfix,showkeys]{revtex4}
\usepackage{graphicx}
\usepackage{epsfig}
\usepackage{amsmath}
\usepackage{amssymb}

\begin{document}
%%%%%%%%%%%%%%%%%% title page information %%%%%%%%%%%%%%%%%%
\title{First-principle calculation of solar cell efficiency under incoherent illumination}

\author{Micha\"{e}l Sarrazin$^{\dagger}$}
\email{michael.sarrazin@unamur.be}
\affiliation{Solid-State Physics Laboratory, Research Center in Physics of Matter and Radiation (PMR), University of Namur, 61 rue de Bruxelles, B-5000 Namur, 
Belgium}

\author{Aline Herman$^{\dagger}$}
\affiliation{Solid-State Physics Laboratory, Research Center in Physics of Matter and Radiation (PMR), University of Namur, 61 rue de Bruxelles, B-5000 Namur, 
Belgium}

\author{Olivier Deparis}
\affiliation{Solid-State Physics Laboratory, Research Center in Physics of Matter and Radiation (PMR), University of Namur, 61 rue de Bruxelles, B-5000 Namur, 
Belgium}

\begin{abstract}Because of the temporal incoherence of sunlight, solar cells efficiency should depend on the degree of coherence of the incident light. However, 
numerical computation methods, which are used to optimize these devices, fundamentally consider fully coherent light. Hereafter, we show
that the incoherent efficiency of solar cells can be easily analytically calculated. The incoherent efficiency is simply derived from the coherent one
thanks to a convolution product with a function characterizing the incoherent light. Our approach is neither heuristic nor empiric but is
deduced from first-principle, i.e. Maxwell's equations. Usually, in order to reproduce the incoherent behavior, statistical methods requiring a high
number of numerical simulations are used. With our method, such approaches are not required. Our results are compared with those from previous works and good 
agreement is found.
\\$^{\dagger}$\textit{These authors have contributed equally to this work.}\end{abstract}


\maketitle

\section{Introduction}

Improvement of solar cell technology as well as cost reduction is an
increasingly challenging topic in the quest of renewable energy sources. One of the
strategy currently followed to reduce the fabrication cost is the use of ultrathin
active layers. This approach could help reducing the cost of photovoltaic
technologies since a smaller quantity of material is used \cite
{Tsakalakos,Nelson}. However, reducing the thickness brings new problems. The fact
that the film thickness could be much lower than the absorption length leads
to a significant reduction of the absorption. Ultrathin technologies need
solutions to keep solar light absorption high. One already well-known
solution is the use of front-side or/and back-side surface texturing to help
coupling incident light into the active layer via light trapping techniques 
\cite{Zeman,Campbell,Yablonovitch,Deparis}. The optimization of
light-trapping structures (see Fig. \ref{fig1}(a)) in solar cell is still of high interest \cite
{Gjessing,Herman}. In this approach, numerical computations are needed to
investigate how electromagnetic field propagates in those devices. Usually,
numerical methods used for calculating the absorption inside a solar
cell (Finite-Difference Time-Domain method (FDTD) \cite{FDTD1,FDTD2},
Rigorous Coupled Wave Analysis (RCWA) \cite{Moharam,code1,code2,code3}, ...) basically
consider the cell response under coherent incident light. Unfortunately, it is
known that the absorption of an optical device depends on the coherent or
incoherent nature of the light. Coherent light leads to oscillations
(such as Fabry-Perot) in the absorption spectrum. Incoherent
light instead, leads to the disappearance of these
oscillations due to destructive-interference effects: a fact that is well known from 
anyone performing measurements in solar cells. Therefore, the
calculated absorption spectrum of the active layer is strongly affected by the
choice of the incident source (coherent or not). Since the absorption
spectrum is used to determine the photocurrent and the efficiency of solar
cells \cite{Herman,Jin,Zhao}, it should be recommended to use only
 incoherent absorption spectrum. Solar light is indeed a strongly
incoherent light source: incoming waves from the sun have a finite
coherence time (finite spectral width). The photocurrent $J$ supplied by
the solar cell is given by: 
\begin{equation}
J=\frac e{hc}\int A(\lambda )S_G(\lambda )\lambda d\lambda  \label{current0}
\end{equation}
where $A(\lambda )$ is the absorption spectrum and $S_G(\lambda )$ the
global power spectral density of the sun. In most studies, the absorption
spectrum $A(\lambda )$ is computed from numerical codes which propagate
the electromagnetic field. Nevertheless, such computed $%
A(\lambda )$ is calculated from coherent fields and then we write $A(\lambda) = A_{coh}(\lambda )$. This
quantity does not correspond to the required effective incoherent absorption $A_{incoh}(\lambda )$
experienced by the solar cell. In order to theoretically predict the performance of
a solar cell, it is therefore very important to improve numerical methods and
to take incoherent incident light into account, i.e. to use $A(\lambda
)=A_{incoh}(\lambda )$ in Eq. (\ref{current0}).

It must be noted that, as far as the propagation of the electric field of incident optical radiation is concerned, a solar cell, whatever the complexity of its 
structure is, behaves as a linear system (thanks to Maxwell equations), which is fully characterized by its scattering matrix (see Appendix \ref{formalism}). 
However, as soon as energy fluxes need to be calculated, linearity does not stand anymore since the intensity (Poynting vector flux) is proportional to
the electric field squared, i.e. $I\propto E\cdot E^*=|E|^2$. The way the square modulus of the fields enters into the calculation of the absorption 
(reflectance, transmittance) of the system is far from being trivial (see, for example, the case of RCWA method, in Appendix \ref{RCWA}). In any case, there 
exists no transfer function relating \textit{linearly} the solar cell absorption to the incident Poynting vector flux. The fact that 
the relevant quantities (absorbed flux, photocurrent,...) do not obey the superposition principle (since $I\propto |E|^2$) prevents us from applying common 
principles of the linear response theory \cite{random_signal}. In particular, the incoherent response cannot be treated in 
the same way as the coherent one, a fact that is known for long time in optics \cite{BornWolf}, at least for simple cases, such as thin films.

In order to overcome this limitation, numerical solutions have been proposed for modeling
incoherent processes \cite
{Prentice_2000,Mitsas,Troparevsky,Niv,Katsidis,Prentice_99,Centurioni,Lee,Santbergen}%
. Basically, light propagation is computed many times for various incident
waves, and the final result relies on a global numerical statistical
analysis. Many recipes and algorithms have been proposed to achieve this
task. Nevertheless, in the present article, we show that it is not necessary to
upgrade numerical codes or to perform time consuming statistical analysis
in order to deal with incoherent light. Indeed, the efficiency of solar
cells under incoherent light illumination can be analytically calculated from the coherent
response computed with usual numerical codes, as explained hereafter. Results
obtained with our method are compared with those from previous studies.

\section{Incoherent absorption}

\begin{figure}[t]
\centerline{\ \includegraphics[width=12 cm]{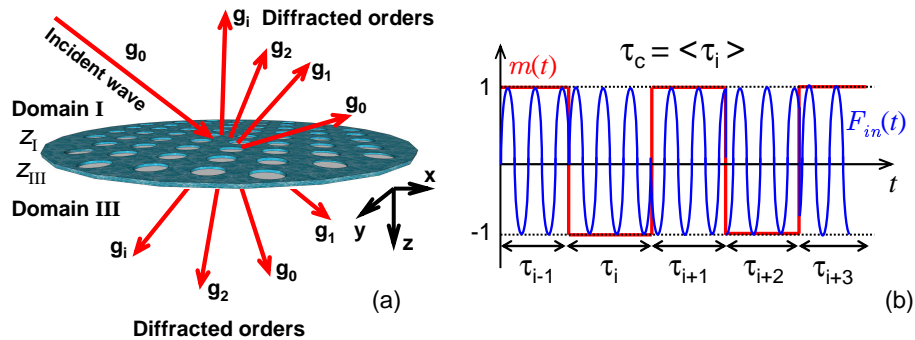}}
\caption{(a) Sketch of a light trapping diffracting structure ($Z_{I}<z<Z_{III}$) 
with many diffraction orders. Domains $I$ and $III$ are respectively the incident and emergence media. (b) Model of field amplitude $F_{in}(t)$ for the incoherent monochromatic incident light. $m(t)$ is the corresponding modulation which characterizes the random phase switching at different times $\tau_{i}$. $\tau_{c}$ is the coherence time, equal to the average value $<\tau_{i}>$.}
\label{fig1}
\end{figure}

When the active domain of a solar cell is illuminated by a coherent
monochromatic light, a steady state is reached in which the intensity of the
electromagnetic field $\left| \mathbf{E}(\omega ,\mathbf{r})\right| ^2$
exhibits a typical stationary pattern. The absorption coefficient $%
A_{coh}(\omega )$ is then given by \cite{BornWolf}: 
\begin{equation}
A_{coh}(\omega )=-\frac{\varepsilon _0\omega }{2P_{in}}\int_V\varepsilon
^{\prime \prime }(\omega ,\mathbf{r})\left| \mathbf{E}(\omega ,\mathbf{r}%
)\right| ^2d^3r  \label{pow}
\end{equation}
where $\varepsilon ^{\prime \prime }(\omega ,\mathbf{r})$ is the imaginary
part of the permittivity of the active medium of the solar cell and $P_{in}$ is the power of the
incident light. $V$ is the volume of the active domain. By contrast, if the light is incoherent, the incident electromagnetic
field exhibits a time-dependent fluctuating behaviour with a characteristic
time $\tau _c$, defined as the coherence time of the incident light.
Therefore, light inside the active volume cannot exhibit the same constructive interference pattern as
in the fully coherent case: incoherence affects the way the light
propagates. In addition, a specific medium mainly interacts with light
through electronic and ionic motions of its components. As a consequence,
the response of a medium to an incident electromagnetic wave is not
instantaneous and occurs with a given response time  $T$ related to
dielectric relaxation time and/or to charge carrier generation and
recombination. If $\tau _c\ll T$, then the medium undergoes many 
different field intensity patterns during the characteristic time $T$. As a
result, the absorption and the generated photocurrent both fluctuate in time.
The effective incoherent absorption $A_{incoh}(\omega )$ (and the
photocurrent) is the time averaged value during a typical time about $T$
(see appendix \ref{formalism}).

It is important to point out that the incoherent absorption $%
A_{incoh}(\omega )$ is physically different from the coherent one $%
A_{coh}(\omega ).$ Indeed, $A_{coh}(\omega )$ can be considered as an
intrinsic property of the solar cell which only depends on its geometry and
on its constitutive materials. By contrast, $A_{incoh}(\omega )$ reflects the effective
measured response of the solar cell while it interacts with its environment,
i.e. for instance when the photocurrent is produced and flows in a closed
electrical circuit.

In a numerical computation approach, the absorption coefficient $A$ can be obtained from
at least two ways. In FDTD for instance, Eq. (\ref{pow}) can be used as such, since maps of
the electromagnetic field are directly computed. In RCWA,
reflection $R$ and transmission $T$ are calculated and $A$ is
deduced from $A=1-R-T$. In the following, we
consider a RCWA formalism of the light propagation (see appendix \ref{RCWA}%
), and we use it to determine the effective incoherent absorption $%
A_{incoh}(\omega )$ (see Appendix \ref{formalism}). The main results of Appendix \ref{formalism}
are summarized hereafter.

Let us consider an incoherent radiation spectrum, the solar spectrum for instance.
Such a spectrum can be considered as constituted by an infinite set of
incoherent quasi-monochromatic spectral lines, each line being expressed by the temporal signal: 
\begin{equation}
\left| F_{in}(t)\right\rangle =m(t)e^{-i\omega _ct}\left|
F_{in}^{(0)}\right\rangle .  \label{iincohT}
\end{equation}
The bracket notation denotes the fact that $\left| F_{in}(t)\right\rangle $
is a supervector which contains the time-dependent complex Fourier components related to
each diffraction orders (see Appendix \ref{RCWA}). $\left|
F_{in}^{(0)}\right\rangle $ describes the incident wave amplitudes, assumed to be non zero only for the 
zeroth diffraction order (see Fig. \ref{fig1}(a) and Appendix \ref{RCWA}).
$\omega _c$ is the angular frequency. The function $m(t)$ is a
modulation which ensures the incoherent behaviour of the spectral line. $%
m(t) $ is a stochastic function which has an average time fluctuation
equal to the coherence time $\tau _c$. As an illustration, a simple but non-restrictive modulation is shown in Fig. \ref{fig1}(b). The
autocorrelation function of a random process has a spectral representation
given by the power spectrum of that process (Wiener Kinchine's theorem) \cite
{random_signal}. The function $m(t)$ can therefore be characterized by its spectral
density: 
\begin{equation}
D(\omega )=\left| m(\omega )\right| ^2  \label{SDT}
\end{equation}
with $m(\omega )=\int m(t)e^{i\omega t}dt$. We then introduce the normalized
incoherence function $I(\omega )$, 
\begin{equation}
I(\omega )=\frac{D(\omega )}{\int D(\omega )d\omega }  \label{normalT}
\end{equation}
which characterizes the incoherence of incident light. Assuming that $m(t)$ describes
a random process which follows a Gaussian distribution, the incoherent
source illuminating the device is characterized by a Gaussian spectral
density $D(\omega )$. As a result, the incoherence function is simply written as: 
\begin{equation}
I(\omega )=\tau _c\sqrt{\frac{\ln 2}{\pi ^3}}e^{-\frac{\ln 2}{\pi ^2}\tau
_c^2\omega ^2}  \label{gaussian}
\end{equation}
with a Full Width at Half Maximum $\Delta \omega =2\pi /\tau _c$. Though the
coherence time is estimated from the solar spectrum, it must be
noted that $D(\omega )$ is not a model of the solar spectrum. $D(\omega )$
simply describes the stochastic behaviour of each spectral line composing 
the whole solar spectrum itself.

It can then be shown (see Appendix \ref{formalism}) that the incoherent
absorption $A_{incoh}(\omega )$ results from the convolution product, noted $\star$, between
the coherent absorption $A_{coh}(\omega )$ and the incoherence function $%
I(\omega )$:

\begin{equation}
A_{incoh}(\omega )=I(\omega )\star A_{coh}(\omega ).  \label{F1}
\end{equation}
This extremely simple formula is easy to use in practice. But its demonstration is not
obvious and is therefore detailed in Appendix \ref{formalism}. Eq. (\ref{F1}) can
also be used to compute incoherent reflection and transmission spectra from
their coherent counterparts (see appendix \ref{formalism}).

\section{Numerical Application}

\begin{figure}[b]
\centerline{\ \includegraphics[width=12 cm]{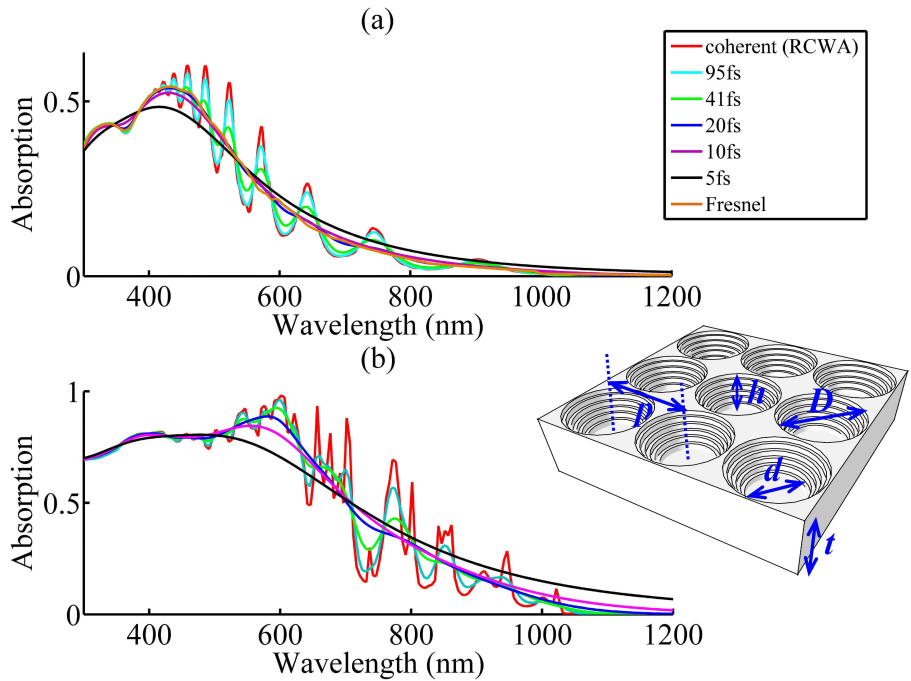}}
\caption{Simulation of the absorption spectra of planar and corrugated $500$
nm-thick c-Si slabs. The coherent spectra were obtained using RCWA and
the incoherent ones using our convolution formula, Eq. (\ref{F1}). 
(a) Absorption spectra of the planar slab according to
various coherence times. (b) Absorption spectra of
the corrugated slab according to various coherence times. Inset: corrugated structure ; $p=500$ nm, $t=500$ nm, $h=300$ nm, $D=450$ nm, $%
d=320 $ nm. The structure follows a super-Gaussian profile with $m=3$ (see Ref. \cite{Herman}).}
\label{Silicon}
\end{figure}

In order to illustrate the usefulness of the convolution formula, i.e. Eq. (\ref{F1}), we use it for
the calculation of the incoherent absorption of a $500$ nm-thick crystalline silicon
(c-Si) slab either planar (see Fig. 2(a)) or corrugated (see Fig. 2(b)).

The first step is the numerical calculation of the coherent absorption
using the RCWA method (see Fig. 2, red lines). In the RCWA
simulations, we use unpolarized light at normal incidence, where the $(x, z)$
plane is the plane of incidence (polar and azimuthal angles equal to $0$)
(see Fig. 1(a)). The permittivities of materials are taken from the literature \cite
{Palik}. The second step is the convolution of the coherent absorption spectrum 
with the incoherence function $I(\omega )$. This step leads to the incoherent absorption spectra.

\begin{figure}[t]
\centerline{\ \includegraphics[width=12 cm]{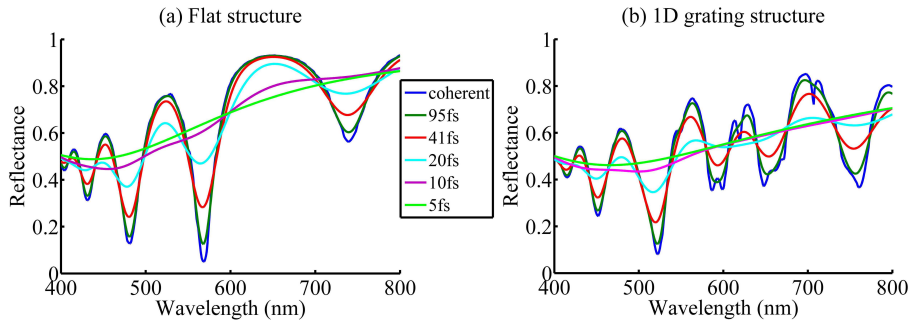}}
\caption{Comparison between the coherent spectra obtained with the RCWA
method (blue lines) and incoherent spectra with various coherence times,
for planar and grating structures, defined in \cite{Lee}.
(a) Reflectance spectrum of an unpatterned c-Si layer ($225$ nm) deposited on
a $75$ nm-thick Au film on a glass substrate. (b) Reflectance spectrum of
the whole grating structure.}
\label{comp_OE}
\end{figure}

The first step is the longest one. For example, using $225$ orders of
diffraction ($15$ plane waves along the $x$ direction and $15$ plane waves
along the $y$ direction), it takes a few hours. The second step is very fast. It
takes only a few minutes on a personal computer. It is an important
improvement in terms of computational time, in comparison with other computational
methods \cite
{Prentice_2000,Mitsas,Troparevsky,Niv,Katsidis,Prentice_99,Centurioni,Lee,Santbergen}%
. In these methods, the first step is performed several times for various
incident waves, and the final result relies on a global numerical statistical
analysis. Five different coherence times are considered here ($95$ fs, $41$ fs, $%
20 $ fs, $10$ fs and $5$ fs). These values are chosen according to the ones
used by W. Lee and coworkers in a previous work \cite{Lee}.

In the case of a planar slab, computed spectra are compared with the approximate 
incoherent absorption spectrum obtained from
a standard analytical expression of the absorption using Fresnel
equations \cite{BornWolf}, but in which propagation phases are roughly set to zero to mimic
incoherence (orange curve in Fig. 2(a)). We note that this kind of analytical expression must be
considered carefully in the present case. Indeed, such an expression
considers the effect of incoherence on light propagation only and does
not consider the incoherent response of the permittivities of the materials,
contrary to our present approach using convolution (see appendix \ref
{formalism}). In spite of that, both spectra (orange and black curves in Fig. 2(a)) 
are quite similar, which validates our
method (in our method, the incoherence limit is approached by taking $\tau_{c}=5$ fs). 
The weak discrepancies are due to the absence of incoherent permittivity response in the Fresnel approach.

We also reproduced some results reported in the literature \cite{Lee}. Two
structures were studied under TM polarization. The first one is a $225$ nm-thick film of crystalline silicon
(c-Si), deposited on a $75$ nm-thick layer of Gold (Au) on a glass substrate.
The second one is a $250$ nm-thick c-Si film deposited on a gold grating on a
glass substrate. The gold grating has a height of $50$ nm, a period of $400$ nm
and a fill factor of $0.5$. The grating is built on a $50$ nm-thick gold
layer. A sketch of both structures can be found in \cite{Lee}. The results
obtained with the convolution formula are presented in Fig. 3 and can be
compared with those shown in figures 2 and 4 in \cite{Lee}.
Some small discrepancies are observed, which are due to the fact
that we did not use the same numerical method (RCWA vs FDTD). However, our
results give the same trends as the results obtained by W. Lee \textit{et al}. It is
therefore possible to account for the light incoherence with a simple method, 
which does not require long computational times contrary to other previously reported methods.

Coming back to Fig. 2, the absorption spectra of a three-dimensional corrugated $%
500$ nm-thick c-Si slab were calculated using the same method as for the planar slab. 
The corrugation follows a super-Gaussian
profile such as the one defined in \cite{Herman} ($m=3$). The period of the
corrugation is $500$ nm and its height is $300$ nm. The absorption spectra
according to various coherence times are shown in Fig. 2(b). We notice
that the value of the coherence time affects the absorption spectrum. The
highest coherence time (here $95$ fs) leads to the strongest oscillations in the spectrum. 
Physically, these oscillations are due to Fabry-Perot resonances (in the thin-slab) 
and to guided mode resonances (enabled by the corrugation).
With the decrease of $\tau_{c}$, these oscillations are expected to be smoothened which 
is actually observed when $\tau_c < 20$ fs. 
This smoothing effect is indeed observed when measuring absorption spectra with an incoherent source.

We finally calculated the photocurent of the corrugated
slab, according to Eq. (\ref{current0}) (see Table \ref{photocurrent}%
). The aim was to quantify the effect of the coherence time on the value of
the photocurrent. The photocurrent fluctuates as the coherence time varies. As expected, the incoherent light behaviour affects both the
absorption spectrum and the photocurrent.

\begin{table}[t]
\begin{center}
\begin{tabular}{ll}
$\tau_{c} (fs)$ & $j_{sc} (mA/cm^2)$ \\ \hline\hline
coherent & 20.69\\ 
95 & 20.70  \\ 
41 & 20.75  \\ 
20  & 20.81 \\ 
10  & 20.93 \\ 
5 & 21.16\\
3 & 21.14\\
2.5 & 21.01
\end{tabular}
\end{center}
\caption{Computed photocurrents related to the corrugated device of Fig. 2(b)
for various coherence times. Photocurrent was integrated over the spectral range: 300 nm - 1200 nm.}
\label{photocurrent}
\end{table}

\section{Conclusion}

We demonstrated that reflection, transmission, and absorption spectra of a
photonic structure illuminated with incoherent light can be easily calculated. The
only input is the knowledge of their coherent counterparts and of the coherence
time of the source (used to determine the incoherence function).
The incoherent response is simply the convolution between a function
accounting for the incoherent source and the coherent response. This result is
theoretically demonstrated from first principle and is confirmed by numerical simulations.
In analogy with signal processing theory, it can be interpreted as the consequence, in the 
frequency domain, of the temporal filtering associated with the intrinsic incoherent modulation.
The proposed method allows to significantly simplify the description of 
incoherent phenomena which are regarded as key problem in solar cells
optimization. The inherent simplicity of our method allows to minimize the computational time
cost and the algorithm complexity. Reflection,
transmission, and absorption spectra, but also the photocurrent, were shown to depend
on the coherence time.

\newpage
\appendix
\numberwithin{equation}{section}

\section{Appendix: Demonstration of the convolution formula}

$\label{formalism}$

\subsection{Scattering matrix as response function}
\label{AA1}

The following demonstration is based on the formalism of the Rigorous
Coupling Wave Analysis (RCWA) \cite{Moharam,code1,code2,code3}. This analysis takes into
account the periodicity of the device and describes the permittivity ($\epsilon$)
through Fourier series. The electromagnetic field is then
described by Bloch waves also expanded in Fourier series. In this formalism,
the Maxwell's equations take the form of a matricial first-order
differential equation in the $z$ variable. The $z$ axis is perpendicular
to the plane ($x$,$y$) where the permittivity is periodic (Fig. 1(a)). The essence
of the method is to solve this equation but is not the topic of the present
article.

Reflected and transmitted field amplitudes are linked to the incident field amplitudes by the
use of the  scattering\ matrix ($S$) which is calculated by solving Maxwell
equations using Fourier series (see Appendix \ref{RCWA}).\ Let us define $%
F_{scat}$ as the scattered field and $F_{in}$ as the incident field, such 
that the associated supervectors are:
\begin{equation}
\left| F_{scat}\right\rangle =\left[ 
\begin{array}{c}
\overline{N}_{III}^{+} \\ 
\overline{X}_{III}^{+} \\ 
\overline{N}_I^{-} \\ 
\overline{X}_I^{-}
\end{array}
\right] ,\left| F_{in}\right\rangle =\left[ 
\begin{array}{c}
\overline{N}_I^{+} \\ 
\overline{X}_I^{+} \\ 
\overline{N}_{III}^{-} \\ 
\overline{X}_{III}^{-}
\end{array}
\right].   \label{incscat}
\end{equation}

The subscripts $I$ and$\ III$ denote the incidence and emergence
media, respectively. The superscripts $+$ and $-$ denote the positive and
negative directions along the $z$ axis associated with the field propagation. All the
Fourier components related to the reciprocal vectors $\mathbf{g}$ are contained in the vector $%
\overline{N}$ or $\overline{X}$ . For each vector $\mathbf{g}$ of the reciprocal
lattice, $N_{I,\mathbf{g}}^{-}$ and $X_{I,\mathbf{g}}^{-}$ are the $s$ and $p
$ polarization amplitudes of the reflected field, respectively. $N_{III,\mathbf{g}}^{+}$
and $X_{III,\mathbf{g}}^{+}$ represent the $s$ and $p$ polarization amplitudes of the
transmitted field. Accordingly, $N_{I,\mathbf{0}}^{+}$ and 
$X_{I,\mathbf{0}}^{+}$ define the $s$ and $p$ polarization amplitudes of
the incident field, respectively. $F_{scat}$\ is connected to $F_{in}$ 
\textit{via} the scattering matrix by: 
\begin{equation}
S(\omega )\left| F_{in}(\omega )\right\rangle =\left| F_{scat}(\omega
)\right\rangle.   \label{S}
\end{equation}
The flux $J$ of the Poynting vector through a unit cell area $\sigma $ for
the incident plane wave in homogeneous medium $I$ is given by: 
\begin{equation}
J_{in}=\sigma \frac 12\varepsilon _0c\sqrt{\varepsilon _I}\cos \theta
\left\langle F_{in}(\omega )\right| \left. F_{in}(\omega )\right\rangle. 
\label{incid}
\end{equation}
For the refracted ($X=R$) or transmitted ($X=T$) wave, it is given by
\begin{equation}
J_X=\left\langle F_{scat}(\omega )\right| C_X^{\dagger }(\omega )C_X(\omega
)\left| F_{scat}(\omega )\right\rangle,   \label{scatt}
\end{equation}
where we define the connection matrices  
\begin{equation}
C_T(\omega )=\left[ 
\begin{array}{cccc}
Q_{III}(\omega ) & 0 & 0 & 0 \\ 
0 & Q_{III}(\omega ) & 0 & 0 \\ 
0 & 0 & 0 & 0 \\ 
0 & 0 & 0 & 0
\end{array}
\right] ,C_R(\omega )=\left[ 
\begin{array}{cccc}
0 & 0 & 0 & 0 \\ 
0 & 0 & 0 & 0 \\ 
0 & 0 & Q_I(\omega ) & 0 \\ 
0 & 0 & 0 & Q_I(\omega )
\end{array}
\right],   \label{matrix}
\end{equation}
with 
\begin{equation}
Q_{I(III)}(\omega )=\left[ 
\begin{array}{ccc}
... & 0 & 0 \\ 
0 & \sqrt{\frac \sigma {2\mu _0\omega }Re\left\{ k_{I(III),\mathbf{g}%
,z}\right\} } & 0 \\ 
0 & 0 & ...
\end{array}
\right].   \label{matrix2}
\end{equation}
In (\ref{matrix2}), $k_{u,\mathbf{g},z}=[(\varepsilon _u(\frac \omega c)^2-|\mathbf{k}_{//}+%
\mathbf{g}|^2)]^{1/2}$ with $\mathbf{k}_{//}$ and $\omega $ being the wave vector
component parallel to the surface and the pulsation of the incident plane
wave, respectively. $\varepsilon _u$\ represents either the
permittivity of the incident medium ($\varepsilon _I$), or of the
emergence medium ($\varepsilon _{III}$). We now introduce a convenient
notation: 
\begin{equation}
J_X=\left\langle F_X(\omega )\right| \left. F_X(\omega )\right\rangle, 
\label{current}
\end{equation}
where 
\begin{equation}
\left| F_X(\omega )\right\rangle =S_X(\omega )\left| F_{in}(\omega
)\right\rangle,   \label{nota}
\end{equation}
with 
\begin{equation}
S_X(\omega )=C_X(\omega )S(\omega ).  \label{nota2}
\end{equation}

\subsection{R, T, A coefficients for a coherent monochromatic incident wave}

$\label{coherent}$

Let us first consider a coherent monochromatic incident wave: 
\begin{equation}
\left| F_{in}(t)\right\rangle =\left| F_{in}^{(0)}\right\rangle e^{-i\omega
_ct}  \label{icoh}
\end{equation}
with an (optical) angular frequency $\omega _c$. The response $\left| F_X(t)\right\rangle $ of
the device is given by \cite{random_signal} 
\begin{equation}
\left| F_X(t)\right\rangle =S_X(t)\star \left| F_{in}(t)\right\rangle
\label{resp1}
\end{equation}
where $\star$ denotes the convolution product. The response can then be written explicitely as: 
\begin{eqnarray}
\left| F_X(t)\right\rangle &=&\int S_X(t-t^{\prime })e^{-i\omega _ct^{\prime
}}dt^{\prime }\left| F_{in}^{(0)}\right\rangle  \nonumber  \label{resp2} \\
&=&e^{-i\omega _ct}S_X(\omega _c)\left| F_{in}^{(0)}\right\rangle,
\label{resp2b}
\end{eqnarray}
where 
\begin{equation}
S_X(\omega _c)=\int S_X(t^{\prime })e^{i\omega _ct^{\prime }}dt^{\prime }
\label{Four1}
\end{equation}
is the Fourier transform of $S_{X}$ scattering matrix defined in (\ref{nota2}). We note that $%
S_{X}(-\omega _c)=S_{X}^{*}(\omega _c)$ since $S_{X}(t)$ must be real.

From (\ref{resp2b}), we set 
\begin{equation}
\left| F_X(t)\right\rangle =\left| F_X^{(0)}\right\rangle e^{-i\omega _ct},
\label{cohresp}
\end{equation}
with 
\begin{equation}
\left| F_X^{(0)}\right\rangle =S_X(\omega _c)\left|
F_{in}^{(0)}\right\rangle.   \label{cohrespb}
\end{equation}
We remind here that $X$ stands either for the reflected ($R$) or the transmitted ($T$) wave.
Reflection $R_{coh}$ and transmission $T_{coh}$ coefficients associated with
a coherent process can be obtained thanks to: 
\begin{equation}
R_{coh}(\omega _c)=\frac{J_R}{J_{in}}=\frac{\left\langle F_R^{(0)}\right|
\left. F_R^{(0)}\right\rangle }{J_{in}}=\frac{\left\langle
F_{in}^{(0)}\right| S_R^{\dagger }(\omega _c)S_R(\omega _c)\left|
F_{in}^{(0)}\right\rangle }{J_{in}},  \label{Rco}
\end{equation}
and 
\begin{equation}
T_{coh}(\omega _c)=\frac{J_T}{J_{in}}=\frac{\left\langle F_T^{(0)}\right|
\left. F_T^{(0)}\right\rangle }{J_{in}}=\frac{\left\langle
F_{in}^{(0)}\right| S_T^{\dagger }(\omega _c)S_T(\omega _c)\left|
F_{in}^{(0)}\right\rangle }{J_{in}}.  \label{Tco}
\end{equation}
The absorption $A_{coh}$ is then simply given by the energy conservation law: i.e. 
$A_{coh}=1-R_{coh}-T_{coh}$. The incident Ponyting flux $%
J_{in}(t)$, i.e. 
\begin{eqnarray}
J_{in}(t) &=&\sigma \frac 12\varepsilon _0c\sqrt{\varepsilon _I}\cos \theta
\left\langle F_{in}(t)\right| \left. F_{in}(t)\right\rangle   \nonumber \\
&=&\sigma \frac 12\varepsilon _0c\sqrt{\varepsilon _I}\cos \theta
\left\langle F_{in}^{(0)}\right| \left. F_{in}^{(0)}\right\rangle, 
\label{Inco}
\end{eqnarray}
turns out to be time independent (see Eq. (\ref{A13}) in Appendix \ref{RCWA}). 
Therefore, the time-averaged incident flux $J_{in}$ is identical to $J_{in}(t)$
\begin{equation}
J_{in}=\frac 1T\int_{-T/2}^{T/2}J_{in}(t)dt=\sigma \frac 12\varepsilon _0c%
\sqrt{\varepsilon _I}\cos \theta \left\langle F_{in}^{(0)}\right| \left.
F_{in}^{(0)}\right\rangle \equiv J_{in}(t).  \label{IncoAv}
\end{equation}
This result is only valid as far as the incident wave is coherent.

\subsection{R, T, A coefficients for an incoherent monochromatic incident wave}

$\label{Incoh1}$

Let us now consider an incoherent quasi-monochromatic incident wave:  
\begin{equation}
\left| F_{in}(t)\right\rangle =\left| F_{in}^{(0)}\right\rangle
m(t)e^{-i\omega _ct} .  \label{iincoh}
\end{equation}
By quasi monochromatic, we mean that the spectral line has a finite though narrow
spectral width.
As explained in the present article, the function $m(t)$ is a modulation which
ensures the incoherent behaviour. On average, $m(t)$ has a coherence time equal 
to $\tau _c$ (see Fig. 1(b)).
Thanks to the Wiener-Khinchine theorem, the autocorrelation function of a
random process has a spectral decomposition given by the power spectrum of
that process\cite{random_signal}. The function $m(t)$ is then equally characterized
by its spectral density: 
\begin{equation}
D(\omega )=\left| m(\omega )\right| ^2,  \label{SD}
\end{equation}
where 
\begin{equation}
m(\omega )=\int m(t)e^{i\omega t}dt  \label{TFm}
\end{equation}
is the Fourier transform of $m(t)$.

The device response is then calculated by the same convolution product as in (\ref{resp1}):
\begin{eqnarray}
\left| F_X(t)\right\rangle &=&S_X(t)\star \left| F_{in}(t)\right\rangle
\label{respin} \\
&=&\left| F_X^{(0)}(t)\right\rangle e^{-i\omega _ct}  \nonumber
\end{eqnarray}
where $X$ again stands for the reflected ($R$) or transmitted ($T$) wave but now

\begin{equation}
\left| F_X^{(0)}(t)\right\rangle =U_X(t)\left| F_{in}^{(0)}\right\rangle,
\label{eff}
\end{equation}
with 
\begin{equation}
U_X(t)=m(t)\star S_X(t)e^{i\omega _ct}.  \label{Ope}
\end{equation}

\subsubsection{Effective response approach}

$\label{Response}$

Upon incoherent illumination, the device with its structure and 
combination of various materials undergoes a set of many incident 
wave trains randomly dephased with respect to each other. These wave trains have the same
pulsation $\omega _c$ and an average duration equal to the coherence time $%
\tau _c$. The coherence time plays a key role while light propagates 
since it prevents constructive interferences from taking place. But it also changes the way 
matter responds to light. Indeed, a specific medium mainly interacts with
light thanks to electronic and ionic motions of its components. As a
consequence, the response of a medium to an incident electromagnetic wave is
not instantaneous and occurs according to the dielectric relaxation time. If
the incoherence time $\tau _c$\ is short enough in comparison with the
relaxation time, the medium response is not coherent. In this case, the
medium response is not simply given by the value of the permittivity at $\omega =\omega _c$.

In this way, we must consider that the full response of the medium
is a time averaged value of the response recorded during a typical time about $T_c$
such that $T_c\gg \tau _c$. The time $T_c$ is the sampling time interval \cite{Lee}
which reflects the non-instantaneity of any measurement process. For instance,
a spectrophotometric measurement is characterized by such a sampling time.
Likewise, the photocurrent of a solar cell is a measure of the response of
the solar cell to the incident light. In this context, the sampling time corresponds to
the recombination/generation time of charged carriers (carrier lifetime).
For instance, in silicon, carrier lifetime ranges from $0.1$ ns to $1$ ms
according to the doping density \cite{Book1}. These delays are very large 
compared to the coherence time of sunlight, which is about $3$ fs \cite
{Book2}. Therefore, the condition $T_c\gg \tau _c$ is well verified in solar cells.

The flux of the Poynting vector for an incoherent incident light $%
J_{X,incoh}$ is given by \cite{Lee}: 
\begin{equation}
J_{X,incoh}=\frac 1{T_c}\int^{T_c}_{0}J_X(t)dt.  \label{Jinc}
\end{equation}
The integral can be easily expressed according to the Fourier transform of $%
J_X(t)$, noted $J_X(\omega)$: 
\begin{equation}
J_{X,incoh}=\int^{\infty}_{-\infty} J_X(\omega )\left( \frac 1{2\pi }\frac{\sin (\omega T_c/2)}{%
\omega T_c/2}\right) d\omega.  \label{inte}
\end{equation}
We note that 
\begin{equation}
\lim_{T_c\rightarrow +\infty }\frac{T_c}{2\pi }\frac{\sin (\omega T_c/2)}{%
\omega T_c/2}=\delta (\omega ),  \label{limit}
\end{equation}
where $\delta$ is the Dirac distribution. Since $T_c\gg \tau _c$, $J_X(\omega )$ should have a
frequency distribution spreading over $\Delta \omega \sim 1/\tau _c$ around zero frequency, quite
comparable to $D(\omega )$, the spectral density of the random process. We can therefore assume that the angular frequencies $%
\omega $ for which $J_X(\omega )$ is significantly different from zero are
such that $T_c\gg 1/\omega $, i.e. $T_c\rightarrow +\infty $ roughly
speaking. We can then fairly consider the following substitution: ($1/2\pi
)\sin (\omega T_c/2)/\left( \omega T_c/2\right) \rightarrow (1/T_c)\delta
(\omega )$. As a result, we get: 
\begin{equation}
J_{X,incoh}\sim \frac 1{T_c}J_X(\omega =0).  \label{Jincfi}
\end{equation}

\subsubsection{Incoherent response}

$\label{Response2}$

Since, using (\ref{eff}), we have ($\dagger$ sign denotes the adjoint matrix, i.e. the conjugate transpose of the matrix) 
\begin{equation}
J_X(t)=\left\langle F_X^{(0)}(t)\right| \left. F_X^{(0)}(t)\right\rangle
=\left\langle F_{in}^{(0)}\right| U_X^{\dagger }(t)U_X(t)\left|
F_{in}^{(0)}\right\rangle,   \label{part}
\end{equation}
we can then deduce from Eqs. (\ref{Ope}), (\ref{Jincfi}) and (\ref{part}): 
\begin{equation}
J_{X,incoh}=\frac 1{T_c}\left\langle F_{in}^{(0)}\right| I_X(\omega
=0)\left| F_{in}^{(0)}\right\rangle   \label{fluxinc}
\end{equation}
where 
\begin{eqnarray}
I_X(\omega ) &=&\int U_X^{\dagger }(t)U_X(t)e^{i\omega t}dt  \nonumber
\label{17} \\
&=&\frac 1{2\pi }U_X^{\dagger }(\omega )\star U_X(\omega ).  \label{Ix}
\end{eqnarray}
From (\ref{Ope}), we deduce that $U_X(\omega )=m(\omega )S_X(\omega +\omega
_c)$ and $U_{X}^{\dagger }(\omega )=m(\omega )S_X^t(\omega -\omega _c)$, where $t$ sign denotes the transpose of the matrix. Then,
(\ref{Ix}) becomes: 
\begin{equation}
I_X(\omega )=\frac 1{2\pi }m(\omega )S_X^{\dagger }(\omega -\omega
_c)\star m(\omega )S_X(\omega +\omega _c),  \label{Ixdev}
\end{equation}
from which, by writing explicitly the convolution product, one deduces 
\begin{eqnarray}
I_X(\omega =0)=\frac 1{2\pi }\int^{\infty}_{-\infty} m(-\omega ^{\prime })S_X^t(-\omega
_c-\omega ^{\prime })m(\omega ^{\prime })S_X(\omega ^{\prime }+\omega
_c)d\omega ^{\prime }.  \label{Ix0}
\end{eqnarray}
Since $m(-\omega ^{\prime })=m^{*}(\omega ^{\prime })$ and $S_X^t(-\omega
_c-\omega ^{\prime })=S_X^{\dagger }(\omega ^{\prime }+\omega _c)$, one
obtains: 
\begin{equation}
I_X(\omega =0)=\frac 1{2\pi }\int^{\infty}_{-\infty} \left| m(\omega _c-\omega ^{\prime
})\right| ^2S_X^{\dagger }(\omega ^{\prime })S_X(\omega ^{\prime })d\omega
^{\prime }.  \label{IxR}
\end{equation}
Then, using Eqs. (\ref{SD}) and (\ref{fluxinc}), we can deduce 
\begin{equation}
J_{X,incoh}=\frac 1{2\pi T_c}\int^{\infty}_{-\infty} D(\omega _c-\omega ^{\prime })\left\langle
F_{in}^{(0)}\right| S_X^{\dagger }(\omega ^{\prime })S_X(\omega ^{\prime
})\left| F_{in}^{(0)}\right\rangle d\omega ^{\prime }.  \label{Jfin}
\end{equation}

\subsubsection{Incident flux}

$\label{Incident}$

Let us estimate the flux  of the incident incoherent wave $J_{in,incoh}$. For
a non dispersive incident medium, we get: 
\begin{eqnarray}
J_{in,incoh}(t) &=&\sigma \frac 12\varepsilon _0c\sqrt{\varepsilon _I}\cos
\theta \left\langle F_{in}(t)\right| \left. F_{in}(t)\right\rangle
\label{Finco} \\
&=&\sigma \frac 12\varepsilon _0c\sqrt{\varepsilon _I}\cos \theta \left|
m(t)\right| ^2\left\langle F_{in}^{(0)}\right| \left.
F_{in}^{(0)}\right\rangle.  \nonumber
\end{eqnarray}
Following the same argument as in section \ref{Response}, at a given
angular frequency $\omega _c$, we assume that the effective flux impinging on the device $%
J_{in,incoh}$ is given by the time average: 
\begin{eqnarray}
J_{in,incoh} &=&\frac 1{T_c}\int^{T_c}_{0}J_{in,incoh}(t)dt  \nonumber \\
&=&J_{in}\frac 1{T_c}\int^{T_c}_{0}\left| m(t)\right| ^2dt  \label{Finco2}
\end{eqnarray}
where $J_{in}$ is the incident flux for the coherent wave, see Eq. (\ref{IncoAv}). Using the same
calculation as in section \ref{Response} for estimating the time
average, one deduces: 
\begin{eqnarray}
\frac 1{T_c}\int^{T_c}_{0}\left| m(t)\right| ^2dt &=&\frac 1{T_c}\frac 1{2\pi
}\left. m^{*}(\omega )\star m(\omega )\right| _{\omega =0}  \nonumber \\
&=&\frac 1{2\pi T_c}\int^{\infty}_{-\infty} D(\omega )d\omega.  \label{aveJ}
\end{eqnarray}
As a consequence, one gets: 
\begin{equation}
J_{in,incoh}=J_{in}\frac 1{2\pi T_c}\int^{\infty}_{-\infty} D(\omega )d\omega.  \label{FincoF}
\end{equation}

\subsubsection{R, T, A coefficients}

$\label{Final}$

Reflection $R(\omega _c)$ and transmission $T(\omega _c)$ coefficients can
always be written as the ratio of a scattered flux to the incident flux, i.e. $%
X_{incoh}(\omega _c)=J_{X,incoh}/J_{in,incoh}$. Then, from (\ref{Jfin})
and (\ref{FincoF}), we can deduce 
\begin{equation}
X_{incoh}(\omega _c)=\frac 1{\int^{\infty}_{-\infty} D(\omega )d\omega }\int D(\omega _c-\omega
^{\prime })\frac{\left\langle F_{in}^{(0)}\right| S_X^{\dagger }(\omega
^{\prime })S_X(\omega ^{\prime })\left| F_{in}^{(0)}\right\rangle }{J_{in}}%
d\omega ^{\prime },  \label{Jfin2}
\end{equation}
 since $J_{in}$ does not depend on $\omega $. For the coherent case,  
we showed previously (\ref{Rco}-\ref{Tco}) that
\begin{equation}
X_{coh}(\omega ^{\prime })=\frac{\left\langle F_{in}^{(0)}\right|
S_X^{\dagger }(\omega ^{\prime })S_X(\omega ^{\prime })\left|
F_{in}^{(0)}\right\rangle }{J_{in}}.  \label{rapp}
\end{equation}
Therefore, the incoherent response can be expressed as a function of the
coherent one: 
\begin{eqnarray}
X_{incoh}(\omega _c) &=&\int^{\infty}_{-\infty} I(\omega _c-\omega ^{\prime })X_{coh}(\omega
^{\prime })d\omega ^{\prime }  \nonumber \\
&=&I(\omega _c)\star X_{coh}(\omega _c)  \label{Final2}
\end{eqnarray}
where 
\begin{equation}
I(\omega )=\frac{D(\omega )}{\int^{\infty}_{-\infty} D(\omega )d\omega }  \label{normal}
\end{equation}
is the normalized spectral density of $m(t)$. As the quantity $X$ stands for 
$R$ and $T$ (reflection and transmission coefficients) and since the
absorption $A$ is defined by $A=1-R-T$, we can deduce the main result 
of our first principle calculation, i.e. Eq. (\ref{F1}),
that is the absorption $A_{incoh}(\omega )$ resulting from an incoherent
scattering process can be related to the same quantity resulting from a
coherent scattering process, i.e. $A_{coh}(\omega )$. The two quantities are related
through a convolution product which involves the normalized spectral density
of $m(t)$, i.e. the incoherence function.

\section{Appendix: Coupled waves analysis (RCWA) formalism}

$\label{RCWA}$

The reader will find more details about the present approach in references \cite{code1,code2,code3}.
We consider as an example a planar dielectric layer with a bidimensional periodic array
described by the dielectric function: 
\begin{equation}
\varepsilon (\mathbf{\rho },\omega )=\sum_{\mathbf{g}}\varepsilon _{\mathbf{g%
}}(\omega )e^{i\mathbf{g\cdot \rho }} . \label{A1}
\end{equation}
In the layer, the dielectric function does not depend on the normal
coordinate $z$, which is used to define the layer thickness, i.e. the layer 
extends from $z=Z_{I}$ to $z=Z_{III}$ (Fig. 1(a)). One sets $\mathbf{\rho }=x_1\mathbf{a}_1+x_2%
\mathbf{a}_2$, according to the unit cell basis ($\mathbf{a}_1$, $\mathbf{a}%
_2)$. In such a system, Bloch's theorem leads to the following
electromagnetic field expression in the layer \cite{code1,code2,code3}:

\begin{equation}
\left[ 
\begin{array}{c}
\mathbf{E} \\ 
\mathbf{H}
\end{array}
\right] =\sum_{\mathbf{g}}\left[ 
\begin{array}{c}
\mathbf{E}_{\mathbf{g}}(z) \\ 
\mathbf{H}_{\mathbf{g}}(z)
\end{array}
\right] e^{i(\mathbf{k+g})\cdot \mathbf{\rho }}e^{-i\omega t}.  \label{A2}
\end{equation}
It can then be easily shown that Maxwell equations can be recasted in the form of a  
first-order differential equation system \cite{code1,code2,code3},
\begin{equation}
\frac d{dz}\left[ 
\begin{array}{c}
\overline{E}_{//}(z) \\ 
\overline{H}_{//}(z)
\end{array}
\right] =\left[ 
\begin{array}{cc}
0 & A \\ 
\widetilde{A} & 0
\end{array}
\right] \left[ 
\begin{array}{c}
\overline{E}_{//}(z) \\ 
\overline{H}_{//}(z)
\end{array}
\right],  \label{A3}
\end{equation}
where $\overline{E}_{//}$ and $\overline{H}_{//}$ are the electric and magnetic field components
parallel to the layer surface.
In the layer, one can then write \cite{code1,code2,code3}: 
\begin{equation}
\left[ 
\begin{array}{c}
\overline{E}_{//}(z_I) \\ 
\overline{H}_{//}(z_I)
\end{array}
\right] =\exp \left\{ \left[ 
\begin{array}{cc}
0 & A \\ 
\widetilde{A} & 0
\end{array}
\right] (z_I-z_{III})\right\} \left[ 
\begin{array}{c}
\overline{E}_{//}(z_{III}) \\ 
\overline{H}_{//}(z_{III})
\end{array}
\right].  \label{A4}
\end{equation}

Let us define the following unit vectors \cite{code1,code2,code3}: 
\begin{equation}
\mathbf{\mu }_{I,\mathbf{g}}=\frac{k_{I,\mathbf{g,}z}}{\sqrt{\varepsilon _I}%
\frac \omega c}\frac{\mathbf{k+g}}{\left| \mathbf{k+g}\right| },  \label{A5}
\end{equation}
\begin{equation}
\mathbf{\eta }_{\mathbf{g}}=\frac{\mathbf{k+g}}{\left| \mathbf{k+g}\right| }%
\times \mathbf{e}_z,  \label{A6}
\end{equation}
\begin{equation}
\mathbf{\chi }_{I,\mathbf{g}}^{\pm }=\mp \,\mathbf{\mu }_{I,\mathbf{g}}+%
\frac{\left| \mathbf{k+g}\right| }{\sqrt{\varepsilon _I}\frac \omega c}%
\mathbf{e}_z.  \label{A7}
\end{equation}
One can then expand the electric and magnetic fields (parallel components) in Fourier series \cite{code1,code2,code3}: 
\begin{eqnarray}
\mathbf{E}_I(\mathbf{\rho },z) &=&\sum_{\mathbf{g}}[N_{I,\mathbf{g}}^{+}%
\mathbf{\eta }_{\mathbf{g}}e^{ik_{I,\mathbf{g},z}(z-z_I)}  \nonumber \\
&&+N_{I,\mathbf{g}}^{-}\mathbf{\eta }_{\mathbf{g}}e^{-ik_{I,\mathbf{g}%
,z}(z-z_I)}  \nonumber \\
&&+X_{I,\mathbf{g}}^{+}\mathbf{\chi }_{I,\mathbf{g}}^{+}e^{ik_{I,\mathbf{g}%
,z}(z-z_I)}  \label{A8} \\
&&+X_{I,\mathbf{g}}^{-}\mathbf{\chi }_{I,\mathbf{g}}^{-}e^{-ik_{I,\mathbf{g}%
,z}(z-z_I)}]e^{i(\mathbf{k+g})\cdot \mathbf{\rho }}  \nonumber  \label{A8b}
\end{eqnarray}
and 
\begin{eqnarray}
\mathbf{H}_I(\mathbf{\rho },z) &=&\frac{\sqrt{\varepsilon _I}}{c\mu _0}\sum_{%
\mathbf{g}}[-N_{I,\mathbf{g}}^{+}\mathbf{\chi }_{I,\mathbf{g}}^{+}e^{ik_{I,%
\mathbf{g},z}(z-z_I)}  \nonumber \\
&&-N_{I,\mathbf{g}}^{-}\mathbf{\chi }_{I,\mathbf{g}}^{-}e^{-ik_{I,\mathbf{g}%
,z}(z-z_I)}  \nonumber \\
&&+X_{I,\mathbf{g}}^{+}\mathbf{\eta }_{\mathbf{g}}e^{ik_{I,\mathbf{g}%
,z}(z-z_I)}  \label{A9} \\
&&+X_{I,\mathbf{g}}^{-}\mathbf{\eta }_{\mathbf{g}}e^{-ik_{I,\mathbf{g}%
,z}(z-z_I)}]e^{i(\mathbf{k+g})\cdot \mathbf{\rho }} .  \nonumber
\end{eqnarray}
The subscripts $I$ and$\ III$ stand for the incident medium and
emergence medium respectively, and the superscripts $+$ and $-$ denote the
positive and negative direction along the $z$ axis for backward ($+$) and forward ($-$) field
propagation. For each vector $\mathbf g$ of the reciprocal lattice, $N_{I,\mathbf{g}%
}^{-}$ and $X_{I,\mathbf{g}}^{-}$ are the $s$ and $p$ polarization amplitudes of the
reflected field, respectively, and $N_{III,\mathbf{g}}^{+}$ and $X_{III,%
\mathbf{g}}^{+}$ , those of the transmitted field. Similarly,
$N_{I,\mathbf{0}}^{+}$ and $X_{I,\mathbf{0}}^{+}$ define the $s$ and $p$
polarization amplitudes of the incident field, respectively.

We can then define a transfer matrix $T$ \cite{code1,code2,code3}:
\begin{equation}
\left[ 
\begin{array}{c}
\overline{N}_I^{+} \\ 
\overline{X}_I^{+} \\ 
\overline{N}_I^{-} \\ 
\overline{X}_I^{-}
\end{array}
\right] =\left[ 
\begin{array}{cc}
T^{++} & T^{+-} \\ 
T^{-+} & T^{--}
\end{array}
\right] \left[ 
\begin{array}{c}
\overline{N}_{III}^{+} \\ 
\overline{X}_{III}^{+} \\ 
\overline{N}_{III}^{-} \\ 
\overline{X}_{III}^{-}
\end{array}
\right].  \label{A10}
\end{equation}
Alternatively, we can express the scattered field against the incident field and
we define a scattering matrix $S$ \cite{code1,code2,code3}: 
\begin{equation}
\left[ 
\begin{array}{c}
\overline{N}_{III}^{+} \\ 
\overline{X}_{III}^{+} \\ 
\overline{N}_I^{-} \\ 
\overline{X}_I^{-}
\end{array}
\right] =\left[ 
\begin{array}{cc}
S^{++} & S^{+-} \\ 
S^{-+} & S^{--}
\end{array}
\right] \left[ 
\begin{array}{c}
\overline{N}_I^{+} \\ 
\overline{X}_I^{+} \\ 
\overline{N}_{III}^{-} \\ 
\overline{X}_{III}^{-}
\end{array}
\right].  \label{A11}
\end{equation}
The flux of the Poynting vector through the unit cell is \cite{code1,code2,code3}: 
\begin{equation}
J=\int_\sigma \frac 12 Re(\mathbf{E}\times \mathbf{H}%
^{*})\cdot \mathbf{e}_z\,dS.  \label{A12}
\end{equation}
We get then \cite{code1,code2,code3}: 
\begin{equation}
J_I^{+}=\frac \sigma {2\mu _0\omega }\sum_{\mathbf{g}}k_{I,\mathbf{g}%
,z}\left[ \left| N_{I,\mathbf{g}}^{+}\right| ^2+\left| X_{I,\mathbf{g}%
}^{+}\right| ^2\right] \Theta (\varepsilon _I(\omega )\frac{\omega ^2}{c^2}%
-\left| \mathbf{k}+\mathbf{g}\right| ^2)  \label{A13}
\end{equation}
for the incident flux in medium $I$,  
\begin{equation}
J_{III}^{+}=\frac \sigma {2\mu _0\omega }\sum_{\mathbf{g}}k_{III,\mathbf{g}%
,z}\left[ \left| N_{III,\mathbf{g}}^{+}\right| ^2+\left| X_{III,\mathbf{g}%
}^{+}\right| ^2\right] \Theta (\varepsilon _{III}(\omega )\frac{\omega ^2}{%
c^2}-\left| \mathbf{k}+\mathbf{g}\right| ^2)  \label{A14}
\end{equation}
for the transmitted flux in medium $III$, and 
\begin{equation}
J_I^{-}=-\frac \sigma {2\mu _0\omega }\sum_{\mathbf{g}}k_{I,\mathbf{g}%
,z}\left[ \left| N_{I,\mathbf{g}}^{-}\right| ^2+\left| X_{I,\mathbf{g}%
}^{-}\right| ^2\right] \Theta (\varepsilon _I(\omega )\frac{\omega ^2}{c^2}%
-\left| \mathbf{k}+\mathbf{g}\right| ^2)  \label{A15}
\end{equation}
for the reflected flux in medium $I$. In (\ref{A13}-\ref{A15}), $\Theta (x)$ is the Heaviside function: i.e.
$\Theta (x)=1$ if $x>0$ and $%
\Theta (x)=0$ otherwise.

\section*{Acknowledgments}
M.S. is supported by the Cleanoptic project (Development of super-hydrophobic anti-reflective coatings for solar glass panels / Convention No.1117317) of the Greenomat program of the Walloon Region (Belgium).
O.D. acknowledges the support of FP7 EU-project No.309127 PhotoNVoltaics (Nanophotonics for ultra-thin crystalline silicon photovoltaics).
This research used resources of the Interuniversity Scientific Computing Facility located at the University of Namur, Belgium, which is supported by the F.R.S.-FNRS under the convention No.2.4617.07.

\end{document}